\documentclass[9pt]{article}
\usepackage[a4paper, bottom=1cm]{geometry}
\usepackage{endnotes}
\usepackage{graphicx}
\usepackage{fullpage}
\usepackage{setspace}
\usepackage{lineno}
\usepackage{url}
\usepackage{subfigure}
\usepackage{epstopdf}
\usepackage{cite}
\usepackage{graphicx}
\usepackage{mdwmath}
\usepackage{mdwtab}
\usepackage{multicol}
\usepackage{amssymb, amsmath}
\usepackage{color}
\usepackage{txfonts}
\usepackage{CJK}
\begin{document}
\title{Description of the Operational Mechanics of a Basel Regulated Banking System}
\author{Jacky Mallett (jacky@iiim.is)}
\maketitle
\section*{Abstract}
This paper presents a description of the mechanical operations of banking
as used in modern banking systems regulated under the Basel Accords, 
in order to provide support for a verifiable and complete description of the 
banking system suitable for computer simulation. Feedback is requested 
on the contents of this document, both with respect to the operations described
here, and any known national, regional or local variations in their structure
and practice.
\section{Introduction}
There appears to be considerable confusion surrounding the precise 
operation of the modern banking system, in particular with respect to the 
regulation of lending and deposit creation, the handling of loan defaults,
and the relationships between 
holdings at the central bank and the bank clearing system and 
the rest of the system.
\par
Simulation of the aggregate behaviour of the banking system is well 
within current computing capabilities, 
and would be highly beneficial both in exploring the impacts of different
regulatory frameworks on the behaviour of the system, and to provide a scientific
foundation for economic understanding of the monetary system.
However for simulation efforts to be successful
an accurate description of the mechanical operations used by banks
in their day to day operations is required and this does not appear
to be currently available either within economic theory, or from 
the regulatory authorities. 
The descriptions that are currently provided by economic textbooks
such as Mankiw\cite{mankiw.1997}, and McConnell\cite{mcconnell.2011}. 
are notably deficient, with important aspects of the system such as 
the precise handling of loan repayments and loan defaults omitted.
\par
This paper aims to provide a clear and verifiable
description of the fundamental operations of the banking system, which can 
then be used to build accurate simulations of its behaviour. We present these
operations following the example of late 19th and early 20th century
bookkeeping manuals on banking such as Shand\cite{shand.1874}, by providing
detailed descriptions of the fundamental bookkeeping operations performed 
by a bank as it
processes deposits, lends money, receives repayment on loans, and handles
loan defaults in a banking system that consists of two banks, A and B, and 
a central bank. 
\section{Double Entry Bookkeeping.}
Banking as we understand it today has emerged over several centuries
from a set of practices first established in Northern Europe by 
medieval goldsmiths and traders\cite{quinn.1994}. It initially developed
as a form of
statistical multiplexing whereby access to physical money in the form of 
gold was managed through day to day bookkeeping practices, operated under the 
assumption that only a fraction of the underlying liabilities (customer 
deposits) would be requested at any one time. Based on this assumption,
goldsmiths would make short term loans of gold to other customers,
and as the chits used to represent gold deposits began to be exchanged directly
a bank based monetary system developed. Over time this system has 
mutated into today's almost entirely electronic transfer based system, however it still
retains the bookkeeping practices of the original system, in particular 
with respect to the relationship between customer deposits, and 
interbank liabilities both in the form of reserves at the central bank, 
and deposits held with other banks in the system. 
The historical antecedents  
of the system are significant, as several of its current features can probably 
only be appreciated within that context.
\par
The mechanical operations used by banks in their day to day processing of money 
and loans, are in large part a creation of the double
entry bookkeeping procedures that evolved to track the customer deposits of physical
money, and the associated lending activities of the banks.  Double entry bookkeeping is based 
on the principle that a general ledger of assets, liabilities and shareholder 
equity is constructed from a series of separate accounts or individual ledger 
books (commonly referred to as T-accounts when presented formally). 
The system of accounts for any bookkeeping entity is deliberately structured so that a
separate and opposite entry must be made into two T-account simultaneously for each action that occurs. 
That is for each debit in one T-Account there must be a separate matching credit in a different T-account, 
and vice versa. The practice was developed by the Florentines in the 13th century, 
initially as an anti-fraud measure, since the separate updates to
two separate books could be structured to require different people to maintain the entries in each book.
\par
In accounting assets are generally the resources owned by a company, and liabilities are 
resources that the bank owes to another separate entity. Customer
deposits at a bank for example, are classified as liabilities, but when physical 
cash is deposited at the bank this is classified as an asset, with the 
corresponding liability being the customer deposit that was created by the deposit of physical money.
The terms debit and credit have very specific meaning within bookkeeping that
are tied to the type of account being operated on. For example, debits 
to accounts classified as dividends, expenses, assets and losses cause the account's balance
to increase, whilst credits to accounts classified as income, revenue, liabilities
and stockholder's equity cause these accounts to be increased. Debits are listed for
all accounts in the left hand column, and credits in the right.
\par
\begin{table}[ht]
\centering
\begin{tabular}{l|rcl|rcl|r}
\multicolumn{2}{c}{Cash} & & \multicolumn{2}{c}{Deposit} & & \multicolumn{2}{c}{Balance}\\
Debit & Credit & & Debit & Credit &  & Asset & Liability \\
\cline{1-2}\cline{4-5}\cline{7-8}
100   &        & &       &  100   &  &  100  &  100      \\
\end{tabular}
\caption{Example of T-Account Cash Handling}
\label{tab:dble_eg}
\end{table}
\par
The balance sheet of assets, versus the liabilities and equity of a bank is built up from the 
set of individual T-accounts.  
In order to maintain this balance, each T-account is classified as either an asset
or a liability. Increases to an asset T-account are then recorded on its debit
side, and decreases as credits; whilst increases to a T-account classified as a
liability are recorded as credits and decreases as debits. Table \ref{tab:dble_eg} shows
an example of this when physical cash 
is deposited at a bank. Two entries are made, a debit into the bank's vault cash account 
which is classified under assets, and a credit into the
customer's deposit account as a liability. The balance of both T-accounts consequently increases,
maintaining equality in the balance book.\footnote{American and English
accounting practices reverse the credit/debit convention, in the English system
increases to an asset account are recorded as a credit. In this document we follow
the American conventions.}
\par
As a consequence the structure surrounding the classification of T-accounts as liabilities or assets
can be somewhat unintuitive. Revenue and capital for example are typically treated as liabilities, 
with the justification
that capital and profits are 'owed' to the shareholders, although more prosaically
this treatment is also required to maintain the balance of bookkeeping operations. Similarly
the handling of loan defaults by banks uses a 'contra-asset' account, which
allows income to be reserved on the Asset side of the ledger against expected
losses. As a result money is removed from the income accounts, that would otherwise be evaluated to 
determine profits and paid as dividends to shareholders.

\section{Bank Model}
Economic models of bank operations are frequently presented at the
annual balance sheet level, following the basic accounting identity:
\begin{equation}
Assets = Liabilities + Stockholder's\:Equity
\label{eqn:basic_actg}
\end{equation}
However, correct analysis of banking behaviour requires a consideration of
the details of monetary flows within the banking system in their 
day to day operations, particularly with respect to the handling of loan 
defaults, which are hidden by this 'identity'. The expanded versions of 
equation \ref{eqn:basic_actg}:
\begin{equation}
Assets = Liabilities + Common\:Stock + Retained\:Earnings
\label{eqn:exp_actg1}
\end{equation}
and 
\begin{equation}
Assets = Liabilities + Common\:Stock + (Income - Expenses) - Dividends\footnote{There are
potential order of evaluation issues with this equation if bracketing is not treated
strictly. It perhaps might also be observed that units are not being correctly treated by
the equality in the equation, and this may cause issues for superficial analyses based on 
it. For example, the majority of assets in the banking context are loans which represent
contractually committed 
flows of money, stock is usually represented in financial instruments that are priced in monetary
units, whilst income, expenses and dividends typically represent money as it is generally 
understood.}
\label{eqn:exp_actg2}
\end{equation}
show the breakdown within the Stockholder's Equity of the bank's day 
to day monetary flows and its capital holdings.
Of particular significance is the definition of expenses, which for a bank
includes its provisions for loan write-offs.
A mistake sometimes found in the economic literature is to simply deduct losses from
Stockholder's Equity in the basic accounting equation \ref{eqn:basic_actg}, rather than
consider the flow implications of the (Income - Expenses) term in
the expanded equation \ref{eqn:exp_actg2}, which indicate that banks can write-off loans 
against income with no effect on stock or capital reserves as long as they remain profitable. 
As stockholder's equity is part of the regulatory
capital for a bank, and in part determines its lending limits, this can lead to incorrect
assumptions about the system's stability.
\par
Analysis of the banking system is further complicated by the increasingly abstract
nature of money, as the banking system continues its evolution away
from physical money to a completely electronic system. The system
was originally based on empirically derived but known ratios between
physical money, the price of precious metals, and the quantity of bank
loans made at each local bank, regulated by the requirement that a fixed percentage of reserves against 
deposits was required to be
held at the central bank.\footnote{The description commonly found in economic textbooks such as 
Mankiw\cite{mankiw.1997}, which appears
to have been derived from the 1931 Macmillan Report to the British Parliament\cite{macmillan.1931}, 
probably authored by Keynes\cite{stamp.1931}, incorrectly shows a reserve being withheld from the 
total customer deposits at the bank, rather
than as additional funds owned by the bank and maintained in a fixed relationship to the quantity
of money represented as deposits.}
While it is not completely correct to equate bank deposits with physical money, if for no other
reason than accounting treatment of the two differs significantly, it is equally
invalid to fail to acknowledge the role bank deposits play as the de facto money
supply in determining the general price level, and indeed have done for over a 
century\cite{redman.1900}.
\par
The original role of bank
deposits as a form of multiplexed access to physical money that is in
day to day use remains embedded in the book keeping accounts, and now creates a feature of
the system that is generally referred to as liquidity, that
is the money available to the bank on the asset side of its balance sheet, to satisfy its 
day to day demands for cash and transfers
within the system through its holdings with the central clearing system and other banks.

\section{Bank Operations}
\subsection{Initial Position}
\begin{table}[ht]
\caption{Initial Position for Bank Operations}
\centering
\begin{tabular}{r | l c l r|r r}
\multicolumn{2}{c}{Central Bank}      &     &            & \multicolumn{2}{c}{Bank A}&          \\
                 Assets & Liabilities &     &            & Assets      & Liabilities &          \\
\cline{1-2} \cline{5-6}
                 400    &             &     & Loans      &   10,000    &   5,000     & Deposit A.C1 \\
                        &             &     &            &             &   5,000     & Deposit A.C2 \\
                        &   200       &     & Reserves   &   200       &             &          \\
                        &             &     & Cash       &   800       &   1,000     & Capital  \\
\cline{5-6}
                        &             &     & Total      &   11,000    &  11,000     &          \\
\\
\\
                        &             &     &            & \multicolumn{2}{c}{Bank B}&          \\
\cline{5-6}
                        &             &     & Loans      &   10,000    &   5,000     & Deposit B.C3 \\
                        &             &     &            &             &   5,000     & Deposit B.C4 \\
                        &  200        &     & Reserves   &   200       &             &          \\
                        &             &     & Cash \& Eq &   800       &   1,000     & Capital  \\
\cline{1-2}\cline{5-6}
                 400    &  400        &     & Total      &   11,000    &  11,000     &          \\
\end{tabular}
\label{tab:start_pos}
\end{table}

Examples in this document are based on a banking system 
consisting of two Banks, A and B, and a simplified Central Bank.  The general ledgers of 
the two banks are shown together with their reserve account 
relationship with the central bank. The other holdings of the central bank are not shown.
The starting position used for the examples in this document is shown in Table \ref{tab:start_pos}.
For the examples shown here, 
the 2\% reserve required of European Banks on accounts with notice periods 
up to 2 years is used. It is assumed all deposit accounts at both banks fall within this 
classification\footnote{Source: 
European Central Bank, \url{http://www.ecb.int/mopo/implement/mr/html/calc.en.html}}. 
Reserve accounts held by banks at the central banks are treated as deposit accounts by the central bank, 
and are consequently classed as liabilities of the central bank. A matching amount of central bank
assets is shown for completeness.
\par
Cash, cash equivalents and reserves represent the bank's own money, its 'liquidity'. Although originally
this would have involved significant holdings of physical cash, today these holdings are
predominantly electronic, and their significance derives from their position in the system
of ledger books in maintaining receipts as funds flow between banks, rather than
directly from customers.  Money paid into or out of the
bank is funnelled through its cash asset journal, with a matching credit or debit
in the account the money is processed for. In an era of electronic operations, this part of the 
bank's operation can be classified as a vestigial structure derived from gold standard era operations, 
but one with significant implications for the behaviour of the larger system.
\subsection{Fundamental Operations}
The following list of bookkeeping operations describe the fundamental mechanical actions
that any bank must perform to maintain its days to day operations.  Potentially some of these actions, 
such as transferring money between accounts can be performed differently when done at the same
bank, than when done between banks as opposed to at the same bank, and consequently
both possibilities are described.
\begin{enumerate}
\item Transfer between accounts at different banks, i.e. cheque or EFT
\item Transfer between accounts at the same bank.
\item Lend money to a customer.
\item Lend money to a customer at a different bank.
\item Borrow from another bank (or central bank)
\item Payment of interest and capital on a bank loan
\item Write off a loan  
\item Increase Capital Holdings
\item Increase Reserve Holdings
\item Central Bank Operations
\begin{itemize}
\item Borrow from the Central Bank (Lender of last resort)
\item Payment of interest on reserves at the Central Bank
\end{itemize}
\end{enumerate}
In the examples below, we first show the set of (credit, debit) tuple operations that
are performed using the American convention (increases in assets are debits), and 
then a worked example following 
the initial position in Table \ref{tab:start_pos}.
\newpage
\subsection{Transfers between Bank Accounts}
\subsubsection{At the same Bank}
When money is transferred between two accounts at the same bank
it is a debit to one account, and a credit to the other, with no
change to the aggregate liability for the bank shown on the balance sheet. 
\par
\begin{table}[ht]
\begin{tabular}{lll}
\multicolumn{2}{l}{Operations} \\
\cline{1-2}
1) & debit customer account (A.C1)     & credit customer account (A.C2)    \\
\end{tabular}
\label{tab:transfer1_op}
\end{table}
\begin{table}[ht]
\centering
\caption{Transfer between accounts at same bank}
\begin{tabular}{r | l c l r|r r}
\multicolumn{2}{c}{Central Bank}      &     &            & \multicolumn{2}{c}{Bank A}&          \\
                 Assets & Liabilities &     &            & Assets      & Liabilities &          \\
\cline{1-2} \cline{5-6}
                 &             &     & Loans      &   10000    	&\color{blue}{4000}   	& Deposit A.C1 \\
                 &             &     &            &           	&\color{blue}{6000}   	& Deposit A.C2 \\
                 &   200       &     & Reserves   &   200     	&           	&          \\
            400  &             &     & Cash \& Eq &   800      	&   1000      	& Capital  \\
\cline{5-6}
                 &             &     & Total      &   11000      	&  11000       	&          \\
\\
\\
                 &             &     &            & \multicolumn{2}{c}{Bank B}&          \\
\cline{5-6}
                 &             &     & Loans      &   10000      	&   5000      	& Deposit B.C3 \\
                 &             &     &            &           	&   5000      	& Deposit B.C4 \\
                 &  200        &     & Reserves   &   200     	&           	&          \\
                 &             &     & Cash \& Eq &   800      	&   1000      	& Capital  \\
\cline{1-2}\cline{5-6}
          400    &  400        &     & Total      &   11000      	&  11000       	&          \\
\end{tabular}
\label{tab:local_transfer}
\end{table}

This is in contrast to the procedure used when money is explicitly transferred between different banks 
shown in section
\ref{sec:transfer2}, which could also be applied to a transfer occurring between customers at the same bank.
While it may seem unlikely that there would be such dramatically different treatment, the potential certainly
appears to exist, and this would have systemic implications if allowed. 
\par
It is also not known what if any differences
in treatment occur when transfers are performed between branches of the same bank. It seems distinctly
possible that both forms of accounting could be in use by different institutions within the same
banking system.\footnote{Banks that 
operate unified bookkeeping across all branches would be able to source larger loans, and could also 
be expected to cause higher monetary expansion rates as they take advantage of a larger liquidity 
channel with the central bank's clearing mechanisms.}
\newpage  
\subsubsection{Transfer between different Banks}
\label{sec:transfer2}
Transfers between the main commercial banks, (clearing banks in the English system) take place through the 
central bank's clearing operations.\footnote{Clearing operations today are usually performed through
a real time transaction based system, but historically depended on an end of day exchange and
balancing approach\cite{campbell.2010}. The exact implementation of the clearing operation,
particularly with respect to its tolerance or otherwise for negative balances during the day, may
have some systemic implications.}
Smaller banks may use accounts at larger banks, rather than direct access to the central
clearing systems. In the example below we will show a transfer through the
reserve accounts held at the central bank. 
\par
For a transfer from customer A.C1 of 1000 at Bank A, to customer B.C3 at Bank B:
\begin{table}[ht]
\begin{tabular}{ll l}
\multicolumn{2}{l}{Operations} \\
\cline{1-2}
1) & debit cash ledger                   & credit reserve ledger            \\
   & credit reserve at central bank      & debit central bank cash account \\
   &                                     &                                 \\
2) & debit  reserve account Bank A       & credit  reserve account bank B   \\
   & debit customer account A.C1         & credit customer account  B.C3   \\
\end{tabular}
\end{table}
\par
The operations are shown in more detail in Tables \ref{tab:transfer1} and \ref{tab:transfer2}, which
show a transfer of 20 from customer A.C1 at Bank A to customer B.C3 at Bank B.

\begin{table}[ht]
\centering
\caption{Transfer: Step 1: Move money to reserves}
\begin{tabular}{r | l c l r|r r}
\multicolumn{2}{c}{Central Bank}      &     &            & \multicolumn{2}{c}{Bank A}&          \\
                 Assets & Liabilities &     &            & Assets      & Liabilities &          \\
\cline{1-2} \cline{5-6}
                 &             &     & Loans      &   10000      	&\color{blue}{5000} & Deposit A.C1 \\
                 &             &     &            &           	        &   5000      	& Deposit A.C2 \\
                 &\color{blue}{220}& & Reserves   &\color{blue}{220}   	&           	&          \\
\color{blue}{420}&             &     & Cash \& Eq &\color{blue}{780}   	&   1000      	& Capital  \\
\cline{5-6}
                 &             &     & Total      &   11000      	&  11000       	&          \\
\\
\\
                 &             &     &            & \multicolumn{2}{c}{Bank B}&          \\
\cline{5-6}
                 &             &     & Loans      &   10000      	&   5000      	& Deposit B.C3 \\
                 &             &     &            &           	&   5000      	& Deposit B.C4 \\
                 &  200        &     & Reserves   &   200     	&           	&          \\
                 &             &     & Cash \& Eq &   800      	&   1000      	& Capital  \\
\cline{1-2}\cline{5-6}
          420    &  420        &     & Total      &   11000      	&  11000       	&          \\
\end{tabular}
\label{tab:transfer1}
\end{table}
\begin{table}[ht]
\centering
\caption{Transfer: Step 2 transfer money to customer A.C1}
\begin{tabular}{r | l c l r|r r}
\multicolumn{2}{c}{Central Bank}      &     &            & \multicolumn{2}{c}{Bank A}&          \\
                 Assets & Liabilities &     &            & Assets      & Liabilities &          \\
\cline{1-2} \cline{5-6}
                 &             &     & Loans      &   10000     &\color{blue}{4980}    	& Deposit A.C1 \\
                 &             &     &            &           	&   5000      	& Deposit A.C2 \\
                 &   200       &     & Reserves   &\color{blue}{200}    	 	&           	&          \\
            420  &             &     & Cash       &   780      	&   1000      	& Capital  \\
\cline{5-6}
                 &             &     & Total      &   10980      	&  10980       	&          \\
\\
\\
                 &             &     &            & \multicolumn{2}{c}{Bank B}&          \\
\cline{5-6}
                 &             &     & Loans      &   10000     &\color{blue}{5020} 	& Deposit B.C3 \\
                 &             &     &            &           	&   5000      	& Deposit B.C4 \\
                 &\color{blue}{220}& & Reserves   &\color{blue}{220}      	&           	&          \\
                 &             &     & Cash \& Eq &   800      	&   1000      	& Capital  \\
\cline{1-2}\cline{5-6}
          420    &  420        &     & Total      &   11020     &  11020       	&          \\
\end{tabular}
\label{tab:transfer2}
\end{table}

\newpage
\subsection{Lending Money}
Similar issues with liquidity considerations and activity that takes place between banks 
as opposed to those at the same bank can be seen with bank lending. Although banks have to
assume that the money they loan may end up on deposit at another bank, and manage their
liquidity exposures appropriately, many banks 
express a clear preference for lending to their own rather than other bank's customers,
a preference that is also recommended in early banking literature. Both alternatives are detailed
below. 
\subsubsection{A loan of money to its own customer.}
Manuals on bank bookkeeping from
the early 20th century indicate that the practice then was to 
enter the loan and the deposit simultaneously in the ledger books as shown here and there is no 
evidence that this practice has ever changed.
\begin{quote}
"If a loan is 
granted, an entry is made in a Customers' loan register, and passed for entry in the Current
Accounts Credit Analysis book. Against the credit so placed to his Current Account, the customer
draws in the ordinary way.

\emph{Bank Bookkeeping and Accounts, Meelboom (p35-p36)\cite{meelboom.1904}}.
\end{quote} 
For a loan of 500  made by Bank A to its own customer A.C1 the operation proceeds as follows:
\begin{table}[ht]
\begin{tabular}{lll}
\multicolumn{2}{l}{Operations} \\
\cline{1-2}
1) & debit loan ledger  & credit customer account (e.g. A.C1)  \\
2) & credit cash ledger & debit  reserve ledger                \\
\end{tabular}
\label{tab:loan1}
\end{table}
\begin{table}[ht]
\centering
\caption{Loan to Bank's own customer}
\begin{tabular}{r | l c l r|r r}
\multicolumn{2}{c}{Central Bank}      &     &            & \multicolumn{2}{c}{Bank A}&          \\
                 Assets & Liabilities &     &            & Assets      & Liabilities &          \\
\cline{1-2} \cline{5-6}
                 &             &     & Loans      &\color{blue}{10500}  &\color{blue}{5500}      	& Deposit A.C1 \\
                 &             &     &            &           		&  5000      	& Deposit A.C2 \\
                 &\color{blue}{210}   & & Reserves&\color{blue}{210}   &           	&          \\
\color{blue}{410}&             &     & Cash       &\color{blue}{790}    &   1000      	& Capital  \\
\cline{5-6}
                 &             &     & Total      &   11500      	&  11500       	&          \\
\\
\\
                 &             &     &            & \multicolumn{2}{c}{Bank B}&          \\
\cline{5-6}
                 &             &     & Loans      &   10000      	&   5000      	& Deposit B.C3 \\
                 &             &     &            &           	&   5000      	& Deposit B.C4 \\
                 &  200        &     & Reserves   &   200     	&           	&          \\
                 &             &     & Cash \& Eq &   800      	&   1000      	& Capital  \\
\cline{1-2}\cline{5-6}
          410    &  410        &     & Total      &   11000      	&  11000       	&          \\
\end{tabular}
\label{tab:loan_cust}
\end{table}
\par
Besides adding to both the loan and customer deposit accounts, the bank may also need to adjust its reserve
provisions with respect to the new level of customer deposits. In this case 10 is transferred
from the bank's cash holdings to the reserve account at the cental bank. We assume in this example
that the bank is still within its risk weighted capital multiple, and does not need to adjust its
capital holdings.

\par
Banks lend money against their asset holdings, with the total 
amount they can lend regulated by reserve requirements at the central bank, capital requirements and in the
case of loans made to customers of other banks, or directly to other banks(interbank lending), 
their cash holdings. 
\par
There are restrictions on the total amount of its loans that a bank can maintain. Under the Basel 
accords, it must be within its risk weighted capital
restrictions, and it must also be able to meet the reserve requirement on its new level of deposits. 
To lend to another bank's customer, the bank must additionally have available liquidity for the transfer
of money for the loan, and in practice since the bank must assume that its funds may be transferred
to other banks, these considerations also apply to loans to its own customers.
\newpage
\subsubsection{Lend to another Bank's Customer}
Lending to a customer at a different bank by contrast requires use of the interbank
transfer mechanisms and follows a different sequence of operations, as shown below.
\par
\begin{table}[ht]
\centering
\begin{tabular}{lll}
\multicolumn{2}{l}{Operations} \\
\cline{1-2}
1) & credit cash holdings at Bank A   & debit reserve holdings at  Bank A   \\
   & credit reserve account at Central Bank & debit assets at Central Bank      \\
2) & credit reserve account at Bank A & debit reserve account at Bank B     \\
   & debit  loan ledger at Bank A     & credit customer account at Bank B   \\
\end{tabular}
\label{tab:loan2}
\end{table}
\par
\begin{table}[ht]
\centering
\caption{Loan to another Bank's Customer}
\begin{tabular}{r | l c l r|r r}
\multicolumn{2}{c}{Central Bank}      &     &            & \multicolumn{2}{c}{Bank A}&          \\
                 Assets & Liabilities &     &            & Assets      & Liabilities &          \\
\cline{1-2} \cline{5-6}
                 &             &     & Loans      &   10000      	&   5000      	& Deposit A.C1 \\
                 &             &     &            &           		&   5000      	& Deposit A.C2 \\
                 &\color{blue}{700} && Reserves   &\color{blue}{700}     	&           	&          \\
\color{blue}{900}   &          &     & Cash \& Eq &\color{blue}{300}  	&   1000      	& Capital  \\
\cline{5-6}
                 &             &     & Total      &   11000      	&  11000       	&          \\
\\
\\
                 &             &     &            & \multicolumn{2}{c}{Bank B}&          \\
\cline{5-6}
                 &             &     & Loans      &   10000      	&   5000      	& Deposit B.C3 \\
                 &             &     &            &           	&   5000      	& Deposit B.C4 \\
                 &  200        &     & Reserves   &   200     	&           	&          \\
                 &             &     & Cash \& Eq &   800      	&   1000      	& Capital  \\
\cline{1-2}\cline{5-6}
          900     &  900       &     & Total      &   11000      	&  11000       	&          \\
\end{tabular}
\label{tab:crossbank1}
\end{table}
\begin{table}[ht]
\centering
\caption{Loan to another Bank's customer}
\begin{tabular}{r | l c l r|r r}
\multicolumn{2}{c}{Central Bank}      &     &            & \multicolumn{2}{c}{Bank A}&          \\
                 Assets & Liabilities &     &            & Assets      & Liabilities &          \\
\cline{1-2} \cline{5-6}
                 &             &     & Loans      &\color{blue}{10500}  &   5000      	& Deposit A.C1 \\
                 &             &     &            &           	&   5000      	& Deposit A.C2 \\
                 &   200       &     & Reserves   &\color{blue}{200}      	&           	&          \\
           400   &             &     & Cash \& Eq &   300      	&   1000      	& Capital  \\
\cline{5-6}
                 &             &     & Total      &   11000    	&  11000       	&          \\
\\
\\
                 &             &     &            & \multicolumn{2}{c}{Bank B}&          \\
\cline{5-6}
                 &             &     & Loans      &   10000    	&\color{blue}{5500}  	& Deposit B.C3 \\
                 &             &     &            &           	&   5000      	& Deposit B.C4 \\
                 &  200        &     & Reserves   &   200     	&           	&          \\
                 &             &     & Cash \& Eq &\color{blue}{1300}     	&   1000      	& Capital  \\
\cline{1-2}\cline{5-6}
          400    & 400         &     & Total      &   11500      	&  11500       	&          \\
\end{tabular}
\label{tab:crossbank2}
\end{table}

This example also illustrates another feature of the system, that the creation of money
in the form of customer deposit entries is independent of the money
on deposit at the central bank (base money) and within the clearing system unless the system
is operating at the limits of its reserve requirements.\footnote{In any banking system where accounts
exist that do not carry reserve requirements (only Net Transaction Accounts require a reserve
in the US system, while time deposits of greater than two years do not require reserves in the
euro-zone), reserve limits effectively only throttle the system's deposit expansion rate, and do
not set absolute limits on expansion.}
\newpage  
\section{Interbank Loan}
A loan to another bank is similar to a loan to a customer at a different bank, with side
effects involving liquidity availability. It is accounted as a liability at the bank receiving
the loan, and as an asset at the bank making it.  
\begin{table}[ht]
\centering
\begin{tabular}{lll}
\multicolumn{2}{l}{Operations} \\
\cline{1-2}
1) & credit cash holdings at Bank A         & debit reserve holdings at Bank A \\
   & credit reserve account at Central Bank & debit assets at Central Bank      \\
2) & credit reserve account at Bank A       & debit reserve account Bank B     \\
   & debit  loan ledger at Bank A           & credit loan liability at Bank B  \\
\end{tabular}
\label{tab:ibl}
\end{table}
\begin{table}[ht]
\centering
\caption{}
\begin{tabular}{r | l c l r|r r}
\multicolumn{2}{c}{Central Bank}      &     &            & \multicolumn{2}{c}{Bank A}&          \\
                 Assets & Liabilities &     &            & Assets      & Liabilities &          \\
\cline{1-2} \cline{5-6}
                 &             &     & Loans      &   10000      	&   5000      	& Deposit A.C1 \\
                 &             &     &            &           	&   5000      	& Deposit A.C2 \\
                 &\color{blue}{700} &&Reserves    &\color{blue}{700}	&           	&          \\
\color{blue}{900}&             &     & Cash \& Eq &\color{blue}{300}&   1000      	& Capital  \\
\cline{5-6}
                 &             &     & Total      &   11000      	&  11000       	&          \\
\\
\\
                 &             &     &            & \multicolumn{2}{c}{Bank B}&          \\
\cline{5-6}
                 &             &     & Loans      &   10000      	&   5000      	& Deposit B.C3 \\
                 &             &     &            &           	&   5000      	& Deposit B.C4 \\
                 &  200        &     & Reserves   &   200     	&           	&          \\
                 &             &     & Cash \& Eq &   800      	&   1000      	& Capital  \\
\cline{1-2}\cline{5-6}
          900     &  900       &     & Total      &   11000    	&  11000       	&          \\
\end{tabular}
\label{tab:ibl1}
\end{table}
\begin{table}[ht]
\centering
\caption{}
\begin{tabular}{r | l c l r|r r}
\multicolumn{2}{c}{Central Bank}      &     &            & \multicolumn{2}{c}{Bank A}&          \\
                 Assets & Liabilities &     &            & Assets      & Liabilities &          \\
\cline{1-2} \cline{5-6}
                 &             &     & Loans      &   10500      	&   5000      	& Deposit A.C1 \\
                 &             &     &            &           		&   5000      	& Deposit A.C2 \\
                 &\color{blue}{200}& & Reserves   &\color{blue}{200}	&           	&          \\
\color{blue}{400}&             &     & Cash \& Eq &\color{blue}{300}    &   1000      	& Capital  \\
\cline{5-6}
                 &             &     & Total      &   11000      	&  11000       	&          \\
\\
\\
                 &             &     &            & \multicolumn{2}{c}{Bank B}&          \\
\cline{5-6}
                 &             &     & Loans      &   10000      	&   5000      	& Deposit B.C3 \\
                 &             &     &            &           		&   5000      	& Deposit B.C4 \\
                 &  200        &     & Reserves   &   200     		&\color{blue}{500}& Loan from Bank A  \\
                 &             &     & Cash \& Eq &\color{blue}{1300}  	&   1000      	& Capital      \\
\cline{1-2}\cline{5-6}
          400    &  400        &     & Total      &   11500      	&  11500       	&              \\
\end{tabular}
\label{tab:ibl2}
\end{table}

\newpage
\section{Loan Repayment}
Loan repayment is broken into two parts, repayment of the principal outstanding on the loan,
and repayment of the interest. Repayment of the principal is a balanced operation, with a simple
deduction from both sides of the balance sheet in the event that the loan is made to a customer
of the same bank which made the loan.  Interest is received as income by the Bank holding the loan,
however its accounting is more complex. Strictly, GAAP requirements are that interest is accrued
on a daily basis, rather than when it is actually paid, but to simplify the presentation this
step is not shown. 
\par
Income is accounted for as a liability, as it nominally represents revenue that will be paid
to the shareholders. In reality, this is a requirement for the double entry bookkeeping operations 
surrounding it to remain balanced, which carries implications for the treatment of expenses,
and in particular loan defaults or write-offs which are treated as an expense in bank accounting. 
Outside the artificial constraints of bookkeeping practices, income received by the bank has
asset like properties, and in particular can be transferred to the asset side when needed
to compensate for loan losses. This element of bank liquidity is systemically interesting
for a number of reasons, in particular loan defaults that can be covered from income do
not impact the quantity of bank lending that is regulated by the capital requirement. Consequently,
when a loan is written off purely against income, the bank is able to extend new loans on its
existing capital base, and subject to liquidity availability the total amount of lending 
it can perform is not affected.
\par
The example shown in Table \ref{tab:interest} and \ref{tab:principal} shows a loan repayment of 100,
 split into two parts, a 40 principal repayment 
and a 60 interest payment made
by customer A.C1 at Bank A. The two payments are processed separately to illustrate the different handling for interest
versus capital repayment, and also that while interest repayment is money supply neutral, principal repayment removes the deposit from the system that was originally created by the loan.
\begin{table}[ht]
\centering
\begin{tabular}{lll}
\multicolumn{2}{l}{Operations} \\
\cline{1-2}
1) & debit principal from customer deposit & credit principal from loan \\
2) & debit interest from customer deposit  & credit interest account for Bank \\
\end{tabular}
\label{tab:loanrepay}
\end{table}
\begin{table}[ht]
\centering
\caption{Principal Repayment}
\begin{tabular}{r | l c l r|r r}
\multicolumn{2}{c}{Central Bank}      &     &            & \multicolumn{2}{c}{Bank A}&          \\
                 Assets & Liabilities &     &            & Assets      & Liabilities &          \\
\cline{1-2} \cline{5-6}
                 &             &     & Loans      &\color{blue}{9960}&\color{blue}{4960}& Deposit A.C1 \\
                 &             &     &            &           	&   5000      	& Deposit A.C2 \\
                 &   200       &     & Reserves   &   200     	&           	&          \\
            400  &             &     & Cash \& Eq &   800      	&   1000      	& Capital  \\
\cline{5-6}
                 &             &     & Total      &   10960      	&  10960       	&          \\
\\
\\
                 &             &     &            & \multicolumn{2}{c}{Bank B}&          \\
\cline{5-6}
                 &             &     & Loans      &   10000      	&   5000      	& Deposit B.C3 \\
                 &             &     &            &           	&   5000      	& Deposit B.C4 \\
                 &  200        &     & Reserves   &   200     	&           	&          \\
                 &             &     & Cash \& Eq &   800      	&   1000      	& Capital  \\
\cline{1-2}\cline{5-6}
          400    &  400        &     & Total      &   11000      	&  11000       	&          \\
\end{tabular}
\label{tab:principal}
\end{table}
\begin{table}[ht]
\centering
\caption{Interest Repayment}
\begin{tabular}{r | l c l r|r r}
\multicolumn{2}{c}{Central Bank}      &     &            & \multicolumn{2}{c}{Bank A}&          \\
                 Assets & Liabilities &     &     & Assets      & Liabilities &          \\
\cline{1-2} \cline{5-6}
                 &             &     & Loans      &   9960    	&\color{blue}{4900}  	& Deposit A.C1 \\
                 &             &     &            &           	&   5000      	 & Deposit A.C2 \\
                 &   200       &     & Reserves   &   200     	&\color{blue}{60}& Interest Income  \\
            400  &             &     & Cash \& Eq &   800      	&   1000      	 & Capital  \\
\cline{5-6}
                 &             &     & Total      &   10960    	&  10960       	 &          \\
\\
\\
                 &             &     &            & \multicolumn{2}{c}{Bank B}&          \\
\cline{5-6}
                 &             &     & Loans      &   10000      	&   5000      	& Deposit B.C3 \\
                 &             &     &            &           	&   5000      	& Deposit B.C4 \\
                 &  200        &     & Reserves   &   200     	&           	&          \\
                 &             &     & Cash \& Eq &   800      	&   1000      	& Capital  \\
\cline{1-2}\cline{5-6}
          400    &  400        &     & Total      &   11000      	&  11000       	&          \\
\end{tabular}
\label{tab:interest}
\end{table}

\newpage
\section{Loan Default} 
Losses on loans are initially treated as an expense for banks,
and are effectively deducted from income, but there are several stages to this process. Additional 
and potentially systemic complexities can occur if
the capital reserve becomes involved. In general loan write-offs are a fairly predictable occurrence, 
several payments have to be missed before a loan can be treated as impaired. Banks
are required to provision against potential losses on a loan at the same time it is made, and 
to continuously monitor and adjust loss provisions to match their anticipated losses. 
Banks also have some freedom 
on how impaired loans are treated, and may elect to write off all or some of the loan, or refinance it.
\par
If loan losses and other expenses significantly exceed income, then
the capital reserve is used to cover the write-off. With respect to the Basel Accord Tier 1 and Tier
2 capital lending provisions, the bank may be over capitalised, in which case there is a buffer
of capital that can be used for this purpose without any impact on its ability to lend with respect to 
its risk weighted capital reserve multiplier. However,
since banks must maintain a limit on their lending that is a multiple of their Tier 1 \& 2 reserve
funds, if losses are sufficiently high they can push the bank out of regulatory
compliance, since it will no longer meet its capital requirement. This last situation is rarely
recoverable without external intervention.
\par
The bookkeeping arrangements that are used to represent the first part of this process
use a contra-asset account for loan losses,
which is then subtracted from gross loans.\footnote{A contra-asset account is an asset account
which has a credit balance, normally asset accounts maintain a debit balance. The contra-asset
account is itself linked to an asset account, and the book value is the net value of the two accounts. 
It is effectively a way of carrying an offsetting allowance forward for loss provisions
or depreciations, linked to specific ledgers for tracking purposes.}
In the example below, we will begin with showing the creation of a loan loss account from income
received, and then assume that its contents are sufficient to cover losses from the loan write down of 
50 for Bank A. 
The significance of the contra-asset designation of the loss provisions account becomes
clearer when the transfer from the interest income account is examined. The funds are
credited to the loss provisions account, but as it is a contra-asset account they act
to reduce the total asset balance. 
\par
When the bank then writes off part of its loan book (50 in the example shown in Table \ref{tab:writeoff3},
the loss provision account is reduced by the amount of the write-off, as is the loan book. Strictly,
the loss provision account is debited, reducing its balance, and the loan account is credited,
also reducing its balance, since it as an Asset account. The net balance of the Assets is unchanged
as a result, since the loss provision account has already accounted for the write off\cite{mecimore.2005}.
\par
As a result, the impact of the write-off on the balance sheet usually precedes the actual write-off.
As long as the loss can be covered from income, then the overall state
of the balance sheet (with respect to the reduction in assets caused by the write-off) can be
restored by new lending. Nor are there any accompanying money supply considerations, since the
money removed from the system by the loan write-off is replaced by the new loan. Some degree of 
loan losses can consequently be absorbed by the system with no systemic repercussions for the 
money supply. There may be implications for the larger economic debt supply, depending on the
subsequent treatment of the loan, which although removed from the bank's balance sheet,
may be sold on for collection. Local practices surrounding the handling of bankruptcy and recourse
and non-recourse lending will play a part there.
\begin{table}[ht]
\centering
\begin{tabular}{lll}
\multicolumn{2}{l}{Operations} \\
\cline{1-2}
1) & debit interest account         & credit loss provisions account \\
2) & credit loan amount             & debit loss provisions account \\
\end{tabular}
\label{tab:loandefault}
\end{table}
\begin{table}[ht]
\centering
\caption{Loan Writeoff - Initial Conditions}
\begin{tabular}{r | l c l r|r r}
\multicolumn{2}{c}{Central Bank}      &     &            & \multicolumn{2}{c}{Bank A}&          \\
                 Assets & Liabilities &     &     & Assets      & Liabilities &          \\
\cline{1-2} \cline{5-6}
                 &             &     & Loans      &   9960      	&   4900      	& Deposit A.C1 \\
                 &             &     &            &           	&   5000      	& Deposit A.C2 \\
                 &             &     &            &           	&     60      	& Interest income \\
                 &   200       &     & Reserves   &   200     	&           	&          \\
            400  &             &     & Cash \& Eq &   800      	&   1000      	& Capital  \\
\cline{5-6}
                 &             &     & Total      &   10960      	&  10960       	&          \\
\\
\\
                 &             &     &            & \multicolumn{2}{c}{Bank B}&          \\
\cline{5-6}
                 &             &     & Loans      &   10000     &   5000      	& Deposit B.C3 \\
                 &             &     &            &           	&   5000      	& Deposit B.C4 \\
                 &  200        &     & Reserves   &   200     	&           	&          \\
                 &             &     & Cash \& Eq &   800      	&   1000      	& Capital  \\
\cline{1-2}\cline{5-6}
          400    &  400        &     & Total      &   11000     &  11000       	&          \\
\end{tabular}
\label{tab:writeoff1}
\end{table}

\begin{table}[ht]
\centering
\caption{Loan Writeoff - Creation of Loss Reserve Account}
\begin{tabular}{r | l c l r|r r}
\multicolumn{2}{c}{Central Bank}      &     &            & \multicolumn{2}{c}{Bank A}&          \\
                 Assets & Liabilities &     &            & Assets      & Liabilities &          \\
\cline{1-2} \cline{5-6}
                 &             &     & Loans      &   9960      &   4900      	& Deposit A.C1 \\
                 &             &     &            &           	&   5000      	& Deposit A.C2 \\
                 &             &     & Loss provision &\color{blue}{(50)}&\color{blue}{10}& Interest income \\
                 &   200       &     & Reserves   &   200     	&           	&          \\
            400  &             &     & Cash \& Eq &   800      	&   1000      	& Capital  \\
\cline{5-6}
                 &             &     & Total      &   10910     &  10910       	&          \\
\\
\\
                 &             &     &            & \multicolumn{2}{c}{Bank B}&          \\
\cline{5-6}
                 &             &     & Loans      &   10000     &   5000      	& Deposit B.C3 \\
                 &             &     &            &           	&   5000      	& Deposit B.C4 \\
                 &  200        &     & Reserves   &   200     	&           	&          \\
                 &             &     & Cash \& Eq &   800      	&   1000      	& Capital  \\
\cline{1-2}\cline{5-6}
          400    &  400        &     & Total      &   11000     &  11000       	&          \\
\end{tabular}
\label{tab:writeoff2}
\end{table}

\begin{table}[ht]
\centering
\caption{Loan Writeoff - Writeoff against loss provisions}
\begin{tabular}{r | l c l r|r r}
\multicolumn{2}{c}{Central Bank}      &     &            & \multicolumn{2}{c}{Bank A}&          \\
                 Assets & Liabilities &     &            & Assets      & Liabilities &          \\
\cline{1-2} \cline{5-6}
                 &             &     & Loans      &\color{blue}{9910}    	&   4900      	& Deposit A.C1 \\
                 &             &     &            &           	&   5000      	& Deposit A.C2 \\
                 &             &     & Loss provision &\color{blue}{(0)}    	&	    10	& Interest income \\
                 &  200        &     & Reserves   &   200      	&           	&          \\
           400   &             &     & Cash \& Eq &   800      	&   1000      	& Capital  \\
\cline{5-6}
                 &             &     & Total      &   10910      	&  10910       	&          \\
\\
\\
                 &             &     &            & \multicolumn{2}{c}{Bank B}&          \\
\cline{5-6}
                 &             &     & Loans      &   10000     &   5000      	& Deposit B.C3 \\
                 &             &     &            &           	&   5000      	& Deposit B.C4 \\
                 & 200         &     & Reserves   &   200     	&           	&          \\
                 &             &     & Cash \& Eq &   800      	&   1000      	& Capital  \\
\cline{1-2}\cline{5-6}
         400     & 400         &     & Total      &   11000     &  11000       	&          \\
\end{tabular}
\label{tab:writeoff3}
\end{table}

\clearpage
\section{Increase Capital}
The capital holdings of a bank are initially the shares purchased by its stockholders when
the bank is founded. The money received by the bank for this purpose becomes its asset
cash holdings. Although the tradable price of shares varies with stock market conditions,
the book value used for common stock held in the bank's capital is the money received
by the bank and initially entered into its cash asset ledger.
\par
Under Basel, capital holdings are divided into two Tiers with regulated definitions for the
financial instruments that can be held in the different tiers, and separate ratios for
the loans that can be extended against their capital holdings. Broadly, Tier 1 consists
of common stock and disclosed reserves or retained earnings. Tier 2 holds undisclosed
reserves, revaluation reserves, additional reserves for loan losses (holdings additional
to the loss provisions described above), and subordinated debt. (Subordinate debt is money
that has been borrowed by the bank, but is subordinate to the claims of the depositors on
bank funds.)
\par
Basel 2 included a Tier 2 category of "hybrid capital instruments" which are financial
instruments having qualities of both debt and equity. The category has proved somewhat
controversial, with a number of such instruments being explicitly forbidden by the regulators,
and appears to be being removed in Basel 3.
\par
There appear
to be no restrictions or controls on increases to the capital reserve, which can be done
from profits, but liquidity would be required for any purchase of financial instruments
such as government treasuries. Since sales of bank stock add to liquidity, this restriction
would not apply to that channel.
\par

the deposit holder of account A.C1.
\begin{table}[ht]
\centering
\begin{tabular}{lll}
\multicolumn{2}{l}{Operations} \\
\cline{1-2}
1) & debit from account A.C1               & debit  reserves at Bank A  \\
   & credit capital                        & credit reserves at Bank B  \\
\end{tabular}
\label{tab:capitalop}
\end{table}
\begin{table}[ht]
\centering
\caption{Sale of Stock to increase Capital}
\begin{tabular}{r | l c l r|r r}
\multicolumn{2}{c}{Central Bank}      &     &            & \multicolumn{2}{c}{Bank A}&         	\\
                 Assets & Liabilities &     &            & Assets      & Liabilities &         	\\
\cline{1-2} \cline{5-6}
                 &             &     & Loans      &   10000    	&\color{blue}{4950}     	& Deposit A.C1   	\\
                 &             &     &            &           	&   5000      	& Deposit A.C2   	\\
                 &\color{blue}{150}      &     & Reserves   &\color{blue}{150}     	&           	&            	\\
            400   &            &     & Cash \& Eq &   800      	&   1000      	& Capital    	\\
\cline{5-6}
                 &             &     & Total      &   10950   	&  10950       	&            	\\
\\
\\
                 &             &     &            & \multicolumn{2}{c}{Bank B}&            	\\
\cline{5-6}
                 &             &     & Loans      &   10000      	&   5000      	& Deposit B.C3   	\\
                 &             &     &            &           	&   5000      	& Deposit B.C4   	\\
                 &\color{blue}{250}       &     & Reserves   &\color{blue}{250}     	&           	&                	\\
                 &             &     & Cash \& Eq &   800    &\color{blue}{1050}    	& Capital        	\\
\cline{1-2}\cline{5-6}
          400     &  400       &     & Total      &   11050      	&  11050       	&                	\\
\end{tabular}
\label{tab:capital_sale}
\end{table}

\newpage
\section{Borrow from Central Bank}
Central bank operations are in principle no different to other bank operations, but operate
from a privileged position in the system with respect to the other banks.
\begin{table}[ht]
\centering
\begin{tabular}{lll}
\multicolumn{2}{l}{Operations} \\
\cline{1-2}
1) & debit Central Bank Assets (loan)      & credit reserve account for Bank A \\
   & debt  reserves at Bank A              & credit loan to Central Bank \\
\end{tabular}
\label{tab:cbloan}
\end{table}
\begin{table}[ht]
\centering
\caption{Loan from Central Bank}
\begin{tabular}{r | l c l r|r r}
\multicolumn{2}{c}{Central Bank}      &     &            & \multicolumn{2}{c}{Bank A}&          \\
                 Assets & Liabilities &     &            & Assets      & Liabilities &          \\
\cline{1-2} \cline{5-6}
                 &             &     & Loans      &   10000      	&   5000      	& Deposit A.C1 \\
                 &             &     &            &           		&   5000      	& Deposit A.C2 \\
                 &\color{blue}{400}& & Reserves   &\color{blue}{400}	&\color{blue}{200}& Loan from Central Bank\\
\color{blue}{600}&             &     & Cash \& Eq &   800               &   1000      	& Capital  \\
\cline{5-6}
                 &             &     & Total      &   11200      	&  11200       	&          \\
\\
\\
                 &             &     &            & \multicolumn{2}{c}{Bank B}&          \\
\cline{5-6}
                 &             &     & Loans      &   10000      	&   5000      	& Deposit B.C3 \\
                 &             &     &            &           		&   5000      	& Deposit B.C4 \\
                 &  200        &     & Reserves   &   200     		&               &              \\
                 &             &     & Cash \& Eq &   800 	   	&   1000      	& Capital      \\
\cline{1-2}\cline{5-6}
          600    &  600        &     & Total      &   11000      	&  11000       	&              \\
\end{tabular}
\label{tab:cbl}
\end{table}

\newpage
\section{Payment of interest on reserve holdings by Central Bank.}
Payment of interest on the reserve holdings is a necessary feature of the system, otherwise
systemic imbalances would result over time from the asymmetric flow within the system
as central bank loans were repaid by the clearing banks. In the example in Table \ref{tab:cb_interest}
it is assumed that 10 has been received by the Central Bank as interest payment on its loans, and
this is now paid to Bank A as interest on its reserves.
\begin{table}[ht]
\centering
\begin{tabular}{lll}
\multicolumn{2}{l}{Operations} \\
\cline{1-2}
1) & debit Central Bank Assets (money)     & credit reserve account at Bank A \\
   & debt  reserves at Bank A              & credit income received at Bank A \\
\end{tabular}
\label{tab:cb_interest}
\end{table}
\begin{table}[ht]
\centering
\caption{Initial Position}                  
\begin{tabular}{r | l c l r|r r}
\multicolumn{2}{c}{Central Bank}      &     &            & \multicolumn{2}{c}{Bank A}&          \\
                 Assets & Liabilities &     &            & Assets      & Liabilities &          \\
\cline{1-2} \cline{5-6}
                 &             &     & Loans      &   10000      	&   5000      	& Deposit A.C1 \\
                 &             &     &            &           		&   5000      	& Deposit A.C2 \\
                 &  200        &     & Reserves   &   200 	        &               &          \\
        410      &   10        &     & Cash \& Eq &   800               &   1000      	& Capital  \\
\cline{5-6}
                 &             &     & Total      &   11000      	&  11000       	&          \\
\\
\\
                 &             &     &            & \multicolumn{2}{c}{Bank B}&          \\
\cline{5-6}
                 &             &     & Loans      &   10000      	&   5000      	& Deposit B.C3 \\
                 &             &     &            &           		&   5000      	& Deposit B.C4 \\
                 &  200        &     & Reserves   &   200     		&               &              \\
                 &             &     & Cash \& Eq &   800 	   	&   1000      	& Capital      \\
\cline{1-2}\cline{5-6}
          410    &  410        &     & Total      &   11000      	&  11000       	&              \\
\end{tabular}
\label{tab:cbi1}
\end{table}
\begin{table}[ht]
\centering
\caption{Payment of Interest on Reserves}
\begin{tabular}{r | l c l r|r r}
\multicolumn{2}{c}{Central Bank}      &     &            & \multicolumn{2}{c}{Bank A}&          \\
                 Assets & Liabilities &     &            & Assets      & Liabilities &          \\
\cline{1-2} \cline{5-6}
                 &             &     & Loans      &   10000      	&   5000      	& Deposit A.C1 \\
                 &             &     &            &           		&   5000      	& Deposit A.C2 \\
                 &\color{blue}{210} && Reserves   &\color{blue}{210}    &\color{blue}{10}& Income  \\
        410      &             &     & Cash \& Eq &   800               &   1000      	& Capital  \\
\cline{5-6}
                 &             &     & Total      &   11000      	&  11000       	&          \\
\\
\\
                 &             &     &            & \multicolumn{2}{c}{Bank B}&          \\
\cline{5-6}
                 &             &     & Loans      &   10000      	&   5000      	& Deposit B.C3 \\
                 &             &     &            &           		&   5000      	& Deposit B.C4 \\
                 &  200        &     & Reserves   &   200     		&               &              \\
                 &             &     & Cash \& Eq &   800 	   	&   1000      	& Capital      \\
\cline{1-2}\cline{5-6}
          410    &  410        &     & Total      &   11000      	&  11000       	&              \\
\end{tabular}
\label{tab:cbi2}
\end{table}

\section{Reserve Holdings}
Two forms of reserve holdings exert regulatory control within the system. 
The capital reserve regulates the total amount of loans that can be made
by the bank, while the reserve held at the central bank, in principle at
least regulates the amount of deposits that the bank may hold. Potentially,
as lending is also linked to deposit creation the central bank reserve
can exert some regulation over lending as well. For this to occur however,
two conditions have to be true. One is that the central bank reserve
requirement is greater than the capital reserve requirement, otherwise
the capital reserve requirement will dominate. The other is that the
reserve requirement is applied to all deposits accounts without exception.
Consequently the regulatory effect  of the central bank reserve can
be diluted in practice. A further consideration, with systemic implications,
is also the mechanisms by which banks are allowed to increase their
central bank reserves. In the USA it seems this can be done through
the deposit of government treasuries, which in practice would remove
systemic control over the quantity of reserves in the system.

\bibliography{finance}
\raggedright
\bibliographystyle{unsrt}

\end{document}